\documentclass[aps,
prd,twocolumn,floatfix,nofootinbib,showpacs,superscriptaddress,tightenlines]{revtex4}

\usepackage{amsmath}
\usepackage{lipsum}
\usepackage{amssymb}
\usepackage{amsthm}
\usepackage{bbold}
\usepackage{dcolumn}
\usepackage{epsfig}
\usepackage{graphics}
\usepackage{graphicx}
\usepackage{longtable}
\usepackage{color}

\usepackage{xspace}
\usepackage{cancel}

\definecolor{darkgreen}{rgb}{0,0.5,0}
\definecolor{purple}{rgb}{0.5,0,0.5}
\definecolor{nblue}{rgb}{0.0,0.0,0.50}
\definecolor{scarlet}{rgb}{1.0,0.2,0}

\newcommand{\be}{\begin{equation}}
\newcommand{\ee}{\end{equation}}
\newcommand{\bea}{\begin{eqnarray}}
\newcommand{\eea}{\end{eqnarray}}
\newcommand{\beas}{\begin{eqnarray*}}
\newcommand{\eeas}{\end{eqnarray*}}
\newcommand{\nn}{\nonumber}

\usepackage[colorlinks=true, pdfstartview=FitV, linkcolor=purple, citecolor= purple, urlcolor=blue]{hyperref}

\newcommand{\sect}[1]{Section~\ref{#1}}

\begin{document}

\title{Constructing Scalar-Photon Three Point Vertex in Massless Quenched Scalar
QED}

\author{L. Albino Fern\'{a}ndez-Rangel}
\author{Adnan Bashir}
\affiliation{Instituto de F\'{i}sica y Matem\'aticas, Universidad
Michoacana de San Nicol\'as de Hidalgo, Apartado Postal 2-82,
Morelia, Michoac\'an 58040, M\'{e}xico.}

\author{L.X. Guti\'{e}rrez-Guerrero}
\affiliation{Departamento de F\'{i}sica, Universidad de Sonora,
Boulevard Luis Encinas J. y Rosales, Colonia Centro, Hermosillo,
Sonora 83000, M\'{e}xico.}

\author{Y. Concha-S\'anchez}
\affiliation{Facultad de Ingenier\'ia Civil, Universidad
Michoacana de San Nicol\'as de Hidalgo, Apartado Postal 2-82,
Morelia, Michoac\'an 58040, M\'{e}xico.}

\date{\today}

\begin{abstract}
Non perturbative studies of Schwinger-Dyson equations (SDEs)
require their infinite, coupled tower to be truncated in order to
reduce them to a practically solvable set. In this connection, a
physically acceptable {\em ansatz} for the three point vertex is
the most favorite choice. Scalar quantum electrodynamics (sQED)
provides a simple and neat platform to address this problem. The
most general form of the three point scalar-photon vertex can be
expressed in terms of only two independent form factors, a
longitudinal and a transverse one. Ball and Chiu have demonstrated
that the longitudinal vertex is fixed by requiring the
Ward-Fradkin-Green-Takahashi identity (WFGTI), while the
transverse vertex remains undetermined. In massless quenched sQED,
we construct the transverse part of the non perturbative
scalar-photon vertex. This construction (i) ensures multiplicative
renormalizability (MR) of the scalar propagator in keeping with
the Landau-Khalatnikov-Fradkin transformations (LKFTs), (ii) has
the same transformation properties as the bare vertex under charge
conjugation, parity and time reversal, (iii) has no kinematic
singularities and (iv) reproduces one loop asymptotic result in
the weak coupling regime of the theory.
\end{abstract}


\maketitle

\date{\today}

\section{Introduction}

Gauge theories of fundamental interactions have been the
cornerstone of describing the physical world at the most basic
level. Their enormous success primarily lies in the region where
the coupling strength is small enough and the tools of
perturbation theory are reliable. However, not all interesting
phenomena can be accessed in this approximation scheme. In the non
perturbative sector of quantum chromodynamics (QCD), two major
phenomena emerge: 1) color confinement, and 2) dynamical chiral
symmetry breaking (DCSB). For studying strongly interacting bound
states, a reliable understanding of these phenomena is essential.
However, it can be achieved solely through non perturbative
techniques such as lattice QCD,
SDEs,~\cite{Dyson:1949ha,Schwinger:1951ex}, chiral perturbation
theory and effective quark models. Keeping this in mind, our
interest is focussed on the study of the physically acceptable
truncations of SDEs beyond perturbation theory.

SDEs are the fundamental equations of motion of any quantum field
theory (QFT). They form an infinite set of coupled integral
equations that relate the $n$-point Green function to the
$(n+1)$-point function. As the simplest example, propagators are
related to the three point vertices, the latter to the four point
functions and so on,~\textit{ad infinitum}. As their derivation
requires no assumption regarding the strength of the interaction,
they are ideally suited for studying interactions like QCD, where
one single theory has diametrically opposed perturbative and non
perturbative facets in the ultraviolet and infrared regimes of
momenta, respectively. Unfortunately, being an infinite set of
coupled equations, they are intractable without some simplifying
assumptions. Typically, in the non perturbative region, SDEs are
truncated at the level of two-point Green functions (propagators).
We must then use an \textit{ansatz} for the full three point
vertex. This has to be done carefully. Otherwise, solutions can be
in conflict with some of the key features of a QFT, such as gauge
invariance of physical observables and renormalizability of the
divergent functions involved, thus jeopardizing the credibility of
the truncation scheme employed.

In contrast with the complicated non abelian scenario of QCD,
quantum electrodynamics (QED) has proved to be a good starting
point in studying the non pertutbative regime of the SDEs. Better
yet, in the absence of Dirac matrices, sQED can offer an even more
attractive model to construct acceptable non perturbative
$ans\ddot{a}tze$ for the vertices involved. In this article, we
set out to construct a scalar-photon three point vertex which must
comply with the following key criteria:

\begin{itemize}

   \item It must satisfy the {\bf Ward-Fradkin-Green-Takahashi
identity}
(WFGTI),~\cite{Ward:1950xp,Green:1953te,Takahashi:1957xn}.

\end{itemize}

Just like in spinor QED and QCD, Ball and
Chiu,~\cite{Ball:1980ay}, provide the non perturbative form of the
longitudinal three point vertex in sQED, which explicitly
satisfies the
WFGTI,~\cite{Ward:1950xp,Green:1953te,Takahashi:1957xn}. We take
it as our starting point.

\begin{itemize}

    \item It must satisfy the {\bf local gauge covariance} properties of the
theory.

\end{itemize}

 Note that although the WFGTI is a consequence
of gauge invariance, it is insufficient to ensure the local gauge
covariance relation of the scalar propagator. In order to ensure
the latter, we demand the transverse part of the vertex to be
constrained by the
LKFTs,~\cite{Landau:1955zz,Fradkin:1955jr,Johnson:1959zz,Zumino:1959wt}.
The LKFTs are a well defined set of transformations which describe
the response of the Green functions to an arbitrary gauge
transformation. These transformations leave the SDEs and the WFGTI
form-invariant and ensure chiral quark condensate is gauge
invariant in spinor QED and QCD, a feature not guaranteed through
satisfying WFGTI alone. Therefore, LKFTs potentially play an
important role in guiding us toward an improved {\em ansatz} for
the three point vertex and imposing gauge invariant chiral
symmetry breaking, see for example
Refs.~\cite{Burden:1993gy,Bashir:2000ur,Bashir:2002sp,Bashir:2004hh,Bashir:2006ga,Bashir:2005wt,Bashir:2008ej,Bashir:2009fv,Aslam:2015nia}.
More recently, these transformations have also been studied in the
world line formalism, where we generalize LKFTs to arbitrary
amplitudes in sQED,~\cite{Ahmadiniaz:2015kfq}.

 The truncation scheme in preserving gauge invariance
of observables has also been studied in simpler gauge theories
such as QED3,
e.g.,~\cite{Maris:1996zg,Bashir:2002dz,Bashir:2002sp,Bashir:2004yt,Fischer:2004nq,Lo:2010fm}.
These works involve introducing constraints of gauge invariance in
the truncations. In Ref.~\cite{Fischer:2004nq}, it was shown that
if one naively employed even the most sophisticated full
Curtis-Pennington (CP) or Ball-Chiu (BC) vertices in different
covariant gauges, they are not sufficient to ensure gauge
invariant results for physical observables and the expected gauge
covariance properties of the fermion propagator. However, in later
articles~\cite{Bashir:2005wt,Bashir:2008fk},  the need to
incorporate the LKFT correctly was emphasized in order to obtain
gauge invariance of corresponding physical observables, such as
the chiral quark condensate and the confinement-deconfinement
phase transition as a function of the number of fermion flavors
($n_f$).

\begin{itemize}

    \item It must ensure the {\bf multiplicative renormalizability} (MR) of the two point
propagator.

\end{itemize}

Studies in massless scalar and spinor QED as well as in QCD,
demonstrate that the LKFT of the wavefunction renormalization
implies an MR form of a power
law,~\cite{Bashir:2004hh,Bashir:2002sp,Bashir:2004mu,Bashir:2006ga,Aslam:2015nia}.
We would like to reiterate that this solution can be reproduced
only with an appropriate choice of the electron-photon three point
vertex, as demonstrated first in Ref.~\cite{Curtis:1990zs}. There
have been a series of works, spanned over a couple of decades,
which construct the electron-photon vertex, implementing the LKFT
and MR of the electron
propagator,~\cite{Curtis:1990zs,Curtis:1991fb,Dong:1994jr,Bashir:1994az,Bashir:1995qr,Bashir:1997qt,Bashir:1999bd,Bashir:2000rv,Kizilersu:2009kg,Bashir:2011ij,Bashir:2011dp}.
In Ref.~\cite{Bashir:2011dp}, MR was implemented for the fermion
propagator and it simultaneously ensures the gauge invariance of
the critical coupling, above which chiral symmetry is dynamically
broken.

In this article, we impose the conditions of MR on the three point
scalar-photon vertex in sQED.  It involves an unknown  function
$W(x)$ of a dimensionless ratio $x$ of momenta, satisfying an
integral constraint which guarantees the MR of the scalar
propagator. In this construction, we assume that the transverse
vertex has no dependence on the angle between the incoming and
outgoing momenta of the scalar particle, an approximation which
can be readily undone through defining an effective transverse
vertex.



\begin{itemize}

    \item It should reduce to its {\bf perturbation theory} Feynman expansion in the limit of
    weak coupling.

\end{itemize}

A truncation of the complete set of SDEs, that maintains gauge
invariance and MR of a gauge theory at every level of
approximation, is perturbation theory. Physically meaningful
solutions of the SDEs must agree with perturbative results in the
weak coupling regime. We use one loop perturbative calculations as
a guiding principle for the three point
vertex,~\cite{Kizilersu:1995iz,Davydychev:2000rt,Bashir:2007qq}.
In our construction in terms of the function $W$ mentioned above,
we explore how perturbation theory provides an additional
constraint. Using a one loop calculation of the scalar-photon
three point vertex presented in
Refs.~\cite{Bashir:2007qq,Bashir:2009xx}, we derive a perturbative
constraint on $W(x)$ to ${\cal{O}}(\alpha)$, in the leading
logarithms approximation (LLA). We ensure that our non
perturbative construction of the said vertex satisfies this
constraint.

\begin{itemize}

    \item It must have the same {\bf symmetry properties} as the bare vertex under
              charge conjugation, parity and time reversal.
    \item One loop perturbation theory suggests that it should be free of any
    {\bf kinematic singularities}. Following Ball and Chiu,~\cite{Ball:1980ay},
    we shall enforce this requirement.

\end{itemize}

The scalar-photon three point vertex $\Gamma^{\mu}(k,p)$ must be
symmetric under the exchange of momenta $k$ and $p$. Moreover, we
do not expect it to have  kinematic singularities as $k^2
\Rightarrow p^2$. We build these features into our construction.

The paper is organized as follows: in~\sect{sec:SDE-SP} we
introduce the SDE for the massless scalar propagator in quenched
sQED. We define the longitudinal and transverse parts of the
scalar-photon vertex and simplify the SDE by performing angular
integration. In~\sect{sec:SP-LKFT}, we study the LKFT for the
scalar propagator to obtain a non perturbative expression for the
wavefunction renormalization which defines this propagator. We
introduce and explain the concept of MR in~\sect{sec:SP-MR}. We
deduce a power law solution for the wavefunction renormalization
of the scalar propagator and compare it with the findings of the
LKFT in~\sect{sec:SP-LKFT}. \sect{sec:Vertex} contains details of
how we impose constraints of the LKFT and MR on the three point
transverse scalar-photon vertex in terms of the function $W(x)$.
In~\sect{sec:PT}, we add additional constraints of one loop
perturbation theory, symmetry properties and the lack of kinematic
singularities. We also construct an explicit example of a non
perturbative massless three point scalar-photon vertex. We present
our conclusions and discussion in~\sect{sec:Conc}.
\newpage

\section{\label{sec:SDE-SP} The SDE for the Scalar Propagator}


The explicit form of the sQED Lagrangian is:
 \bea
   {\cal L}_{\rm sQED} &=& -\frac{1}{4} F_{\mu \nu} F^{\mu \nu} -
   \frac{1}{2 \xi} \, \left( \partial^{\mu} A_{\mu} \right)^2 +
   \left( \partial^{\mu} \varphi^* \right)
   \left( \partial_{\mu} \varphi \right) \nn \\
   &-& m^2 \varphi^* \varphi -i e \left( \varphi^* \partial^{\mu} \varphi -
    \varphi \partial^{\mu} \varphi^*  \right) A_{\mu} \nn \\
   &+& 2 e^2
    \varphi^* A^{\mu} \varphi A_{\mu} - \frac{\lambda}{4} (\varphi^*
    \varphi)^2 \,.
 \eea
The detailed derivation of the SDEs for relevant Green functions
for this sQED Lagrangian already exists in
literature,~\cite{Binosi:2006da}. The SDE for the scalar
propagator $S(k)$, in the quenched approximation, is shown in
Fig.~\ref{ScalarPropagatorSDE}:
\begin{figure}[ht]
\includegraphics[width=0.45\textwidth]{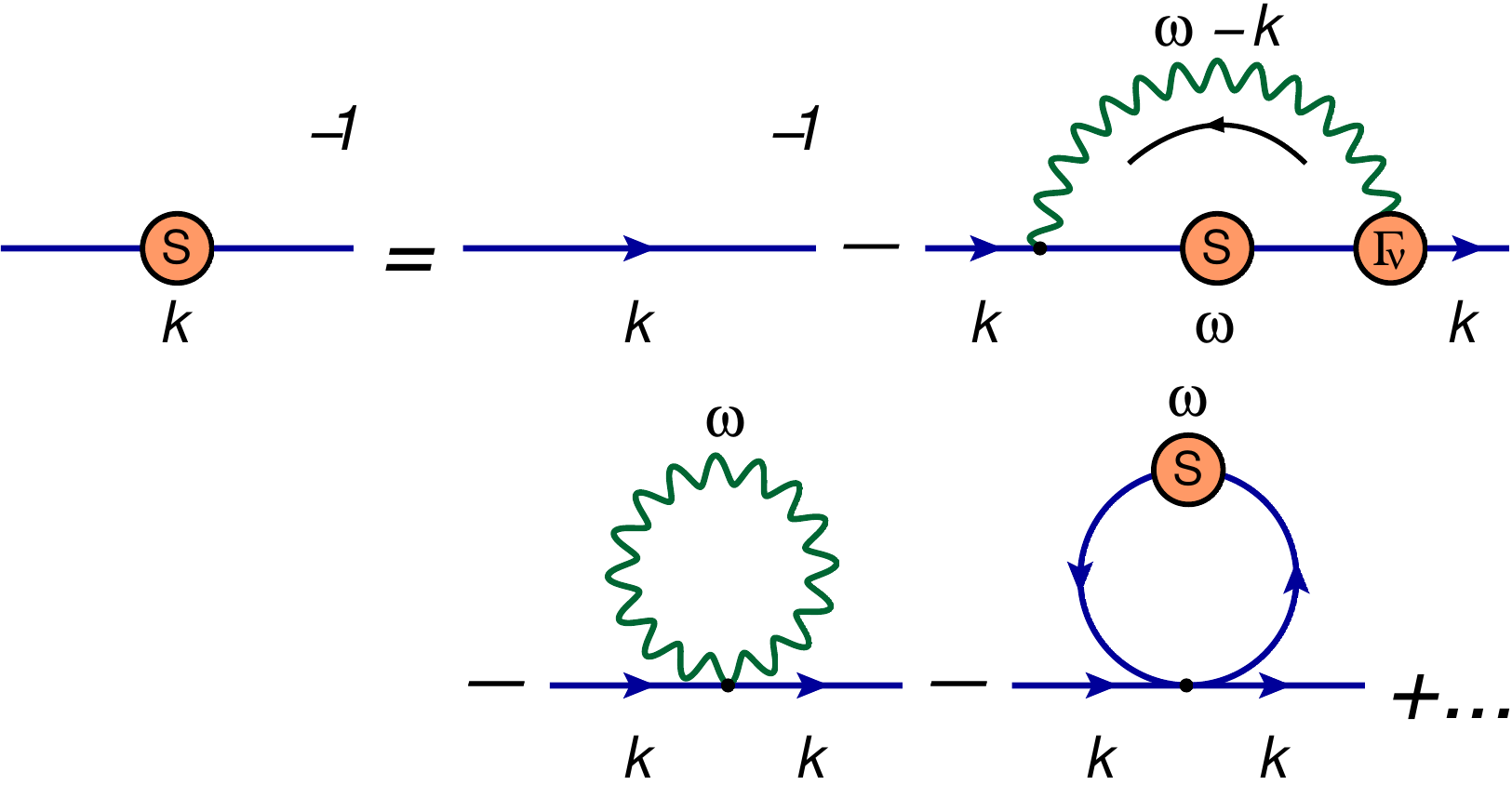}
 \caption{The SDE for the scalar
propagator. The color-filled solid blobs labelled with $S$ and
$\Gamma_{\nu}$ stand for the  full scalar propagator and the full
scalar-photon vertex, respectively. The dots ($\cdots$) represent
all the diagrams whose contribution begins at the two loop level.}
\label{ScalarPropagatorSDE}
\end{figure}

\noindent Mathematically, this is written as:
\begin{eqnarray}
-iS^{-1}(k) & = & -iS_{0}^{-1}(k) \nonumber \\
&+& e^{2} \int_{M}{ \frac{d^{4} \omega}{(2\pi)^{4}} (\omega +
k)^{\mu} S(\omega)
\Gamma^{\nu}(\omega,k) \Delta_{\mu\nu}(q) } \nonumber \\
&-& e^{2} \int_{M}{ \frac{d^{4} \omega}{(2\pi)^{4}}
\Gamma^{\mu\nu}_0 (k,-\omega,k,\omega) \Delta_{\mu\nu}(\omega) }
\nonumber \\
&-& \int_{M}{ \frac{d^{4} \omega}{(2\pi)^{4}} S(\omega)
\Gamma_0(k,\omega) } + \cdots \,, \label{gap equation}
\end{eqnarray}
where $e$ is the electromagnetic coupling, $q=\omega - k$, and the
subscript $M$ indicates integration over the entire Minkowski
space.
$\Delta_{\mu\nu}(\omega)$ and $S_{0}(k)$ are the bare photon and
scalar propagators. $S(k)$ is the full scalar propagator. For
massless scalars, $S(k)$ can be expressed in terms of the
so-called wavefunction renormalization $F(k^{2},\Lambda^{2})$, so
that
\begin{equation}
S(k)=\frac{F(k^{2},\Lambda^{2})}{k^{2}} \,, \label{scalar
propagator}
\end{equation}
where $\Lambda$ is the ultraviolet cut-off used to regularize the
divergent integrals involved. The bare scalar propagator is given
by $S_{0}(k)=1/k^{2}$. The bare photon propagator is
\begin{equation}
\Delta_{\mu\nu}(q)=\frac{1}{q^{2}} \left[ g_{\mu\nu} +(\xi-1)
\frac{q_{\mu}q_{\nu}}{q^{2}} \right] \,, \label{photon propagator}
\end{equation}
and it remains unrenormalized in the quenched approximation.
 $\Gamma^{\mu\nu}_0 (k,-\omega,k,\omega)=2 i e^2
g^{\mu \nu}$ and $\Gamma_0(k,\omega) = - i \lambda$ are the bare
four point scalar-scalar-photon-photon and the four-scalar
vertices, respectively. The last two diagrams of the gap equation,
Eq.~(\ref{gap equation}), will be referred to as the photon and
the scalar bubble diagrams, in that order.
$\Gamma^{\nu}(\omega,k)$ is the full three point scalar-photon
vertex, for which we must make an \textit{ansatz} in order to
solve Eq.~(\ref{gap equation}). The WFGTI for this vertex, i.e.,
\begin{equation}
q_{\mu}\Gamma^{\mu}(\omega,k)=S^{-1}(\omega)-S^{-1}(k) \,,
\label{WGTI for the 3-point vertex}
\end{equation}
allows us to decompose it as a sum of longitudinal and transverse
components, as suggested by Ball and Chiu,~\cite{Ball:1980ay}:
\begin{equation}
\Gamma^{\mu}(\omega,k)=\Gamma_{L}^{\mu}(\omega,k)+\Gamma_{T}^{\mu}(\omega,k)
\,. \label{Ball-Chiu vertex decomposition}
\end{equation}
The \textit{longitudinal} part $\Gamma_{L}^{\mu}(\omega,k)$
satisfies the WFGTI, Eq.~(\ref{WGTI for the 3-point vertex}), by
itself, and the \textit{transverse} part
$\Gamma_{T}^{\mu}(\omega,k)$, which remains completely
undetermined, is naturally constrained by
\begin{equation}
q_{\mu}\Gamma_{T}^{\mu}(\omega,k)=0 \,. \label{transverse part
definition}
\end{equation}
Moreover,
\begin{equation}
\Gamma_{T}^{\mu}(k,k)=0 \,. \label{kklimit}
\end{equation}
In order to satisfy the WFGTI in a manner free of kinematic
singularities, we follow Ball and Chiu and write
\begin{equation}
\Gamma_{L}^{\mu}(\omega,k) = \frac{S^{-1} (\omega)-S^{-1}(k) }{
\omega^{2} - k^{2} } (\omega + k)^{\mu} \,. \label{Longitudinal
vertex}
\end{equation}
This construction implies that the ultraviolet divergences solely
reside in the longitudinal part. Moreover, recall the following
relations between the renormalized and bare quantities:
 \begin{eqnarray}
  S^{R}(p) = {\cal Z}_2^{-1} S(p) \,, \;
 \Gamma^{\mu}_{R}(k,p) = {\cal Z}_1  \Gamma^{\mu}(k,p) \,.
 \end{eqnarray}
Thus, the form of the longitudinal vertex in
Eq.~(\ref{Longitudinal vertex}) guarantees the relation ${\cal
Z}_1 = {\cal Z}_2$. Consequently, the running of the coupling is
dictated by the corrections to the photon propagator alone. In the
approximation of quenched sQED, the coupling does not run. If we
unquench the theory, it is easy to calculate ${\cal Z}_3$ and the
running coupling constant with the well known expression:
 \begin{eqnarray}
  \alpha(Q^2) = \frac{\alpha(Q_0^2)}{1 - \left( \alpha(Q_0^2)/12 \pi \right)
  {\rm ln}(Q^2/Q_0^2) } \,.
 \end{eqnarray}
The ultraviolet finite transverse vertex can be expanded out in
terms of one unknown function
$\tau(\omega^{2},k^{2},q^{2})$,~\cite{Ball:1980ay}:
\begin{equation}
\Gamma_{T}^{\mu}(\omega,k) = \tau(\omega^{2},k^{2},q^{2})
T^{\mu}(\omega,k) \,, \label{Transverse vertex}
\end{equation}
where
\begin{equation}
T^{\mu}(\omega,k) = \left( \omega \cdot q \right) k^{\mu} - \left(
k \cdot q \right) \omega^{\mu} \label{Tensor_mu}
\end{equation}
is the transverse basis vector in the Minkowski space and fulfils
Eqs.~(\ref{transverse part definition},\ref{kklimit}). To begin
with, the form factor $\tau(\omega^{2},k^{2},q^{2})$ is an
unconstrained scalar function (representing an $8$-fold
simplification of the spinor QED/QCD case).

Following the non perturbative vertex construction/truncation of
Refs.~\cite{Ball:1980ay,Bashir:2007qq}, our analysis ensures that
gauge invariance (in terms of the WFGTI) for the scalar propagator
and the scalar-photon vertex is satisfied. Within our truncation,
another source of gauge non-invariance in the scalar propagator
could be the lack of implementation of LKFT, a feature of the bare
as well as BC vertices. We make sure that our {\em ansatz} for the
transverse part satisfies this constraint non perturbatively.
Photon propagator also has its Ward identity but we work
throughout in the quenched approximation. Therefore, within the
confines of our assumptions, it receives no corrections and hence
the four point diagrams we have discarded do not affect the
correct gauge invariance properties of the scalar propagator. They
will be essential, for example, in ensuring the transversality of
the photon propagator in unquenched sQED, not investigated in the
present work.
 Furthermore, there are also restrictions of the gauge
transformations on how the three point vertex is related to the
four point vertex, constraining the form of the latter. In Ref.
~\cite{Bashir:2009xx}, two of the present authors exploited these
constraints to carry out its non perturbative construction
consistent with WFGTI which relates three point vertices to the
four point ones. There is an undetermined part which is transverse
to one or both the external photons, and needs to be evaluated
through perturbation theory. They present in detail how the
transverse part at the one loop order can be evaluated for
completely general kinematics of momenta involved in covariant
gauges and dimensions. In this article, our focus is on
constraining the non perturbative three point scalar-photon
vertex, capturing its key features, in particular its gauge
covariance properties, its perturbative expansion in the LLA as
well as the MR of the scalar propagator.

We make use of Eqs.~(\ref{WGTI for the 3-point vertex},
\ref{Ball-Chiu vertex decomposition},\ref{Longitudinal vertex},
\ref{Transverse vertex},\ref{Tensor_mu}) in the gap equation,
i.e., Eq.~(\ref{gap equation}), and then Wick rotate it to the
Euclidean space to write:
\begin{eqnarray}
\hspace{-1cm} \frac{1}{F(k^{2},\Lambda^{2})} & = &
1-\frac{\alpha}{4\pi^{3}} \frac{1}{k^{2}} \int_{E}{ d^{4}\omega
\frac{1}{q^{2}} \left\{ \left[ 1-\frac{S(\omega)}{S(k)} \right]
 \right. }  \nonumber \\
& & \hspace{-1cm} \times \left[1 + (\xi -1)
\frac{\omega^{2}-k^{2}}{q^{2}}
+2\frac{k^{2}}{\omega^{2}-k^{2}} \right. \nonumber \\
& & \hspace{-1cm} \left. \left. +2\frac{\omega \cdot
k}{\omega^{2}-k^{2}} \right]   - 2S(\omega)
\tau(\omega^{2},k^{2},q^{2}) \Delta^{2} \right\}  \,,
\label{equation for F without assumptions}
\end{eqnarray}
where $\Delta^2=\left( \omega \cdot k \right)^{2} - \omega^{2}
k^{2}$, $\alpha=e^{2}/4\pi$ is the bare coupling constant, and the
subscript $E$ indicates integration over the whole Euclidean
space. Note that we have neglected the photon and the scalar
bubble diagrams as well as the diagrams whose contribution begins
at the two loops level, since they do not contribute to leading
logs in the one loop calculation, as we shall discuss later. At
this stage, it appears impossible to proceed any further because
of the dependence of $\tau$ on the angle between the incoming and
outgoing momenta $\omega$ and $k$ of the scalar particle. We shall
assume that the transverse vertex has no dependence on this angle,
i.e., it is independent of $q^{2}$. Consequently, this vertex is
only an effective one which will allow us to capture many key
features of the theory in a simple manner.
 This assumption allows us to carry out the angular
integration in Eq.~(\ref{equation for F without assumptions}). In
this sense, we are calculating an \textit{effective} transverse
vertex. Note that it is easy to undo this independent angle
approximation exactly. This has been explained and employed in
Refs.~\cite{Bashir:1997qt,Bashir:2011vg} for the case of spinor
QED. Based upon the results found in these articles and our
cross-check for sQED, we conclude that the qualitative
implications of the {\em ansatz} of the three point scalar-photon
vertex are insignificant, and hence we do not report corresponding
findings.

The angular integration leads us to
\begin{eqnarray}
\frac{1}{F(k^{2},\Lambda^{2})} & = & 1 - \frac{\alpha}{4\pi}
\int_{0}^{ k^{2}}{ d\omega^{2} \frac{\omega^{2}}{k^{2}} \left[
1-\frac{S(\omega)}{S(k)} \right] } \nonumber \\
& & \hspace{1cm} \times \left[ \frac{(2 - \xi)}{k^{2}} +
\frac{1}{\omega^{2}-k^{2}} \left( 2 + \frac{\omega^{2}}{k^{2}}
\right) \right] \nonumber \\
& & \hspace{-.5cm} - \frac{\alpha}{4\pi} \int_{
k^{2}}^{\Lambda^{2}}{ d\omega^{2} \left[ 1-\frac{S(\omega)}{S(k)}
\right] \left\{ \frac{3}{\omega^{2} - k^{2}} + \frac{\xi}{k^{2}}
\right\} }
\nonumber \\
& & \hspace{-.5cm} + \frac{\alpha}{8\pi} \int_{0}^{ k^{2}}{
d\omega^{2} \omega^{2} S(\omega) \tau(\omega^{2},k^{2}) \left(
\frac{\omega^{4}}{k^{4}} - 3 \frac{\omega^{2}}{k^{2}} \right) }
\nonumber \\
& & \hspace{-.5cm} + \frac{\alpha}{8\pi} \int_{
k^{2}}^{\Lambda^{2}}{ d\omega^{2} \omega^{2} S(\omega)
\tau(\omega^{2},k^{2}) \left(
\frac{k^{2}}{\omega^{2}} - 3 \right) } \,. \nonumber \\
& & \label{equation for F with effective vertex}
\end{eqnarray}
At this point, it is obvious that we require the knowledge of the
form factor $\tau(\omega^{2},k^{2})$ to find the wavefunction
renormalization $F(k^{2},\Lambda^{2})$. However, this problem can
be inverted. The requirements of LKFT and the MR of
$F(k^{2},\Lambda^{2})$ can tightly constrain the function
$\tau(\omega^{2},k^{2})$. We would like to stress that these
constraints will be valid only within our truncation scheme which
consists of the set of assumptions and hypotheses we have detailed
before. We study them in the next section.

\section{\label{sec:SP-LKFT} Scalar Propagator and LKFT}

These transformations have the simplest structure in the Euclidean
coordinate space. Therefore, we start by defining the Fourier
transformations between the scalar propagators in coordinate and
momentum spaces:
 \bea
 {\cal S}_E(x;\xi)&=& \int \frac{d^dk}{(2\pi)^d} \ {\rm e}^{-i\mathbf{k} \cdot \mathbf{x}} \, S_E(k;\xi)\,,\label{f2x}\\
 S_E(k;\xi)&=& \int d^dx \ {\rm e}^{i\mathbf{k} \cdot
 \mathbf{x}} \, {\cal S}_E(x;\xi)\;. \label{f2p}
 \eea
Notice a slight modification of notation that we shall use in this
section: $S(p) \Rightarrow S(p;\xi)$ for the sake of clarity.
Moreover, we use the notation ${\cal S}$ for the propagator in the
coordinate space in order to specify that its functional
dependence is different from that of $S$, the same propagator in
the momentum space. The subscript $E$ stands for the Euclidean
space.

 The LKFT relating the coordinate space scalar propagator in a given gauge
$\xi_0$ to the one in an arbitrary covariant gauge $\xi$ reads:
 \bea {\cal S}_E^{\rm LKFT}(x;\xi) =
 {\cal S}_E(x;\xi_0){\rm e}^{-i [\Delta(0)-\Delta(x)]} \;,\label{LKprop}
 \eea
 where
 \bea
 \Delta (x)&=&-i (\xi-\xi_0) e^2 (\mu x)^{4-d} \int \frac{d^dk}{(2\pi)^d} \frac{{\rm e}^{-i\mathbf{k} \cdot \mathbf{x}}}{k^4}\nn\\
 &=&-\frac{i (\xi-\xi_0) e^2}{16 (\pi)^{d/2}} (\mu x)^{4-d} \Gamma\left(\frac{d}{2}-2\right) \;.\label{deltad}
 \eea
Here, $\mu$ is a mass scale introduced for dimensional purposes;
it ensures that in every dimension $d$, the coupling $e$ is
dimensionless. For the four dimensional case, we expand around
$d=4- 2 \epsilon$ and use
 \bea
 \Gamma \left( - {\epsilon} \right) &=& -
 \frac{1}{\epsilon} - \gamma + {\cal O}(\epsilon) \;, \nn \\
 x^{\epsilon} &=& 1 + \epsilon {\rm ln} x + {\cal O}(\epsilon^2) \; .
 \eea
Therefore,
 \bea
 \Delta(x)=i\frac{ (\xi-\xi_0)
e^2}{16\pi^{2-\epsilon}}\Big[\frac{1}{\epsilon}+\gamma+2\ln \mu
x+\mathcal{O}\,(\epsilon)\Big]\,.
 \eea
Note that in the term proportional to $\ln x$, one cannot simply
put $x=0$. Therefore, we need to introduce a cutoff scale
$x_{\min}$. We then arrive at
 \bea
 \Delta(x_{\rm min})-\Delta(x)=-i\ln \left(\frac{x^2}{x_{\rm
 min}^2}\right)^\nu\,,
 \eea
 with
$\nu={\alpha (\xi-\xi_0)}/{(4\pi)}$. If we have the knowledge of
the propagator in one gauge, we can transform it to any other
gauge dictated by the LKFT:
 \bea
 {\cal S}_E^{\rm LKFT}(x;\xi) &=& {\cal S}_E(x;\xi_0)\,
 {\rm e}^{-i\big(\Delta(x_{\min})-\Delta(x)\big)} \nn \\
 &=& {\cal S}_E(x;\xi_0)\,
 \Big(\frac{x^2}{x_{\rm min}^2}\Big)^{-\nu}\,.
 \label{propagator}
 \eea
 Let us start from the tree level massive scalar propagator
 \bea
 S_E(k;\xi_0) = - \frac{1}{k^2 +m^2} \,.
 \eea
 Its Fourier transformation into the coordinate space is:
 \bea
  {\cal S}_E(x;\xi_0) = -\frac{m}{4\pi^2x} K_1(mx)\;,
 \eea
 where $K_1(mx)$ is the modified Bessel function of the second kind.
 The LKFT readily yields:
 \bea
  {\cal S}_E^{\rm LKFT}(x;\xi) &=& -\frac{m}{4\pi^2x}
K_1(mx)\left(\frac{x^2}{x^2_{\rm min}} \right)^{-\nu}\;.
 \eea
 We can Fourier transform this result back to the momentum space
 to get
 \bea
 S_E^{\rm LKFT}(k;\xi) &=& -\frac{1}{m^2}\left(
\frac{m^2}{ -\Lambda^2}\right)^{\nu} \Gamma(1-\nu) \Gamma(2-\nu)
\nn
\\
 &&   \times  \; _2F_1\left(1-\nu,2-\nu;2;-\frac{k^2}{m^2}\right)
 \,,
 \eea
where we have made the identification $4/x_{\rm min}^2 \rightarrow
-\Lambda^2$. This is the non perturbative LKFT expression for the
scalar propagator, starting from its knowledge at the tree level
in the gauge $\xi_0$. To evaluate it in the massless limit, we
make use of the identity
 \bea
 {}_2F_1 \left(a,b;c;z  \right) = (1-z)^{-a} {}_2F_1 \left(a,c-b;c;\frac{z}{z-1}\right)
 \eea
to rewrite the scalar propagator as follows:
 \bea
 && \hspace{-1.2cm} S_E^{\rm LKFT}(k;\xi)=  - \left( \frac{1}{-\Lambda^2} \right)^{\nu} \Gamma(1-\nu) \Gamma(2-\nu)
   \nn \\
 && \hspace{-.5cm} \times \,(k^2 + m^2)^{\nu-1} \,  _2F_1\left(1-\nu,\nu;2;-\frac{k^2}{k^2 + m^2}\right) \,.
 \eea
The massless limit now yields
 \bea
  S_E^{\rm LKFT}(k;\xi)&=& - \frac{1}{k^2} \;
  \frac{\Gamma(1-\nu)}{\Gamma(1+\nu)}\; \left( -\frac{k^2}{\Lambda^2}
  \right)^{\nu} \,.
  \eea
 This is a power law with exponent $\nu$. Expanding it out in the
 powers of coupling, retaining the leading logarithms and writing the result in the Minkowski space, we get:
 \bea
  S^{\rm LKFT}(k;\xi)&=&  \frac{1}{k^2} \left[ 1 + \, \frac{\alpha (\xi - \xi_0)}{4 \pi} \; {\rm ln}
  \left(  \frac{k^2}{\Lambda^2} \right) \right] \,. \label{Prop-LKFT}
  \eea
 It implies
 \bea
  F^{\rm LKFT}(k^{2},\Lambda^{2}) &=&   1 + \, \frac{\alpha (\xi - \xi_0)}{4 \pi} \; {\rm ln}
  \left(  \frac{k^2}{\Lambda^2} \right) \,. \label{WFR-LKFT}
 \eea
 This result provides constraints on the transverse scalar-photon vertex
 through Eq.~(\ref{equation for F with effective vertex}). Before
 we set about exploiting this constraint, we would like to connect
 Eq.~(\ref{WFR-LKFT}) with perturbation theory and MR of the scalar propagator in the next section.

%
%

\section{\label{sec:SP-MR} Scalar Propagator and MR}

MR of the scalar propagator requires the renormalized $F_{R}$ to
be related to the unrenormalized $F$ through a multiplicative
factor ${\cal Z}_{2}$ by
\begin{equation}
F_{R}(k^{2},\mu^{2}) = {\cal Z}_{2}^{-1}(\mu^{2},\Lambda^{2})
F(k^{2},\Lambda^{2}) \,, \label{Scalar propagatos MR eq}
\end{equation}
where $\mu$ plays the role of an arbitrary renormalization scale.
Within a truncation scheme which focusses only on logarithmic
divergences, it is possible to write down the above functions as
perturbative series involving terms of the form $\alpha^{n}
\ln^{n}$ (the so called leading log terms).  We should keep in
mind that the sQED has features of a $\varphi^4$ scalar field
theory as well as the spinor QED. It has both quadratic and
logarithmic ultraviolet divergences. Our truncation scheme makes
it resemble the spinor QED or QCD, problems of our eventual
interest.

In the LLA, we then have
\begin{eqnarray}
F(k^{2},\Lambda^{2}) &=& 1 + \sum_{n=1}^{\infty} \alpha^{n} A_{n}
\ln^{n} \left( \frac{k^{2}}{\Lambda^{2}} \right) \,, \label{F unrenormalized expansion} \\
{\cal Z}_{2}^{-1}(\mu^{2},\Lambda^{2}) &=&  1 +
\sum_{n=1}^{\infty} \alpha^{n} B_{n} \ln ^{n} \left(
\frac{\mu^{2}}{\Lambda^{2}} \right)
\,, \label{Z function} \\
F_{R}(k^{2},\mu^{2}) &=& 1 + \sum_{n=1}^{\infty} \alpha^{n} C_{n}
\ln ^{n} \left( \frac{k^{2}}{\mu^{2}} \right) \,. \label{F
renormalized expansion}
\end{eqnarray}
(Note that the next to leading logs (NLL) are of the type
$\alpha^n \ln^{n-1}$ and so on.) The MR condition,
Eq.~(\ref{Scalar propagatos MR eq}), requires
\begin{equation}
A_{n}=C_{n}=(-1)^{n}B_{n}=\frac{A_{1}^{n}}{n!}\,,
\label{coefficients}
\end{equation}
so that the functions $F$, $F_{R}$ and ${\cal Z}_{2}^{-1}$ obey a
power law behavior. Thus the non perturbative solution of
Eq.~(\ref{Scalar propagatos MR eq}) for $F$ in the LLA is
\begin{equation}
F(k^{2},\Lambda^{2}) = \left( \frac{k^{2}}{\Lambda^{2}} \right)
^{\beta} \,, \label{F unrenormalized lead log expansion}
\end{equation}
where the anomalous dimension $\beta$ is unknown at the non
pertubative level. This is in contrast with perturbation theory,
where $\beta=\alpha A_{1}$ is obvious from
Eq.~(\ref{coefficients}). It is straightforward to calculate
$A_{1}$ in one loop perturbation theory: taking the tree level
values $\Gamma^{\nu}(\omega,k)=(\omega + k)^{\nu}$ and
$S(\omega)=1/\omega^{2}$ in the gap equation, i.e., Eq.~(\ref{gap
equation}), we get, on Wick rotating it to  the Euclidean space,
\begin{eqnarray}
\frac{1}{F(k^{2},\Lambda^{2})} &=& 1 - \frac{\alpha}{4\pi^{3}}
\frac{1}{k^{2}} \int_{E}{ \frac{d^{4} \omega}{\omega^{2}}
\frac{(\omega + k)^{2}}{q^{2} } } \nonumber \\
& & \hspace{-.7cm} - \frac{\alpha}{4\pi^{3}} \frac{(\xi -
1)}{k^{2}} \int_{E}{ \frac{d^{4} \omega}{\omega^{2}}
\frac{(\omega^{2} - k^{2})^{2}}{q^{4}} } \,. \label{one-loop F}
\end{eqnarray}
Note that we have dropped the photon and the scalar bubble
contributions as they do not contribute to the LLA. Angular
integration of Eq.~(\ref{one-loop F}) yields
\begin{eqnarray}
\frac{1}{F(k^{2},\Lambda^{2})} &=& 1 + \frac{\alpha (\xi -
3)}{4\pi} \int_{k^{2}}^{\Lambda^{2}}{ \hspace{-.15cm}
\frac{d\omega^{2}}{\omega^{2}} }
\nonumber \\
& & \hspace{-1.5cm} + \frac{\alpha}{4\pi} \frac{(\xi - 3)}{k^{4}}
\int_{0}^{k^{2}}{ \hspace{-.3cm} d\omega^{2} \, \omega^{2} } -
\frac{\alpha}{4\pi} \frac{\xi}{k^{2}} \int_{0}^{\Lambda^{2}}{
\hspace{-.3cm} d\omega^{2}} \,. \label{one-loop F 2}
\end{eqnarray}
After carrying out the radial integration in the above
Eq.~(\ref{one-loop F 2}), dropping the quadratic and quartic
divergencies ($\Lambda^{2}$ and $\Lambda^{4}$) coming from the
last two terms on the right hand side of Eq.~(\ref{one-loop F 2})
and conserving only the logarithmic divergence (as we are
interested in the LLA), we have
\begin{eqnarray}
F(k^{2},\Lambda^{2}) &=& 1+ \frac{\alpha (\xi - 3)}{4\pi} \ln
\left( \frac{k^{2}}{\Lambda^{2}} \right) \,. \label{one-loop
sacalar propagator}
\end{eqnarray}
Comparing Eqs.~(\ref{WFR-LKFT},\ref{one-loop sacalar propagator}),
we deduce that $\xi_0=3$ is the correct choice for sQED till one
loop order in perturbation theory. This is unlike the case of
spinor QED, where Landau gauge $\xi=0$ works well for the same
order of approximation.

Comparing expression~(\ref{one-loop sacalar propagator}) with the
perturbative expansion~(\ref{F unrenormalized expansion}) to
one-loop order, we see that $A_{1}= (\xi - 3)/4\pi$. Therefore,
perturbation theory suggests that the anomalous dimension
in~(\ref{F unrenormalized lead log expansion}) is
\begin{equation}
\beta = \frac{\alpha (\xi - 3)}{4\pi} \,,  \label{anomalous
dimension}
\end{equation}
see
also~\cite{Delbourgo:1977vh,Delbourgo:2003wd,Kreimer:2004xz,Bashir:2007qq}.
One can readily note that the power behavior of~(\ref{F
unrenormalized lead log expansion}), with $\beta$ given
in~(\ref{anomalous dimension}), is the solution of the following
integral equation:
\begin{equation}
\frac{1}{F(k^{2},\Lambda^{2})} = 1 + \frac{\alpha (\xi - 3)}{4\pi}
\int_{k^{2}}^{\Lambda^{2}}{ \frac{d \omega^{2}}{\omega^{2}}
\frac{F(\omega^{2},\Lambda^{2})}{F(k^{2},\Lambda^{2})} } \,.
\label{integral equation for F}
\end{equation}
This term can be separated out in Eq.~(\ref{equation for F with
effective vertex}) to impose the required condition of MR on the
transverse form factor $\tau(\omega^2,k^2)$. This is what we study
in the next section.

\section{\label{sec:Vertex} The Transverse Vertex}

Eq.~(\ref{integral equation for F}) imposes the following
restriction on the transverse vertex through Eq.~(\ref{equation
for F with effective vertex}):
\begin{eqnarray}
 -2 \int_{0}^{k^{2}}{ \hspace{-.3cm} d \omega^{2}
\left\{ \frac{3}{k^{2}} + \frac{(3-\xi)}{k^{2}}
\frac{\omega^{2}}{k^{2}}
+\frac{3}{\omega^{2} - k^{2}} \right. } & & \nonumber \\
\left. +\frac{(\xi - 3)}{k^{2}}
\frac{F(\omega^{2},\Lambda^{2})}{F(k^{2},\Lambda^{2})}
-\frac{3}{\omega^{2} - k^{2}}
\frac{F(\omega^{2},\Lambda^{2})}{F(k^{2},\Lambda^{2})} \right\}
& & \nonumber \\
-2 \int_{k^{2}}^{\Lambda^{2}}{ \hspace{-.3cm} d \omega^{2} \left\{
\frac{3}{\omega^{2} - k^{2}} -\frac{3}{\omega^{2} - k^{2}}
\frac{F(\omega^{2},\Lambda^{2})}{F(k^{2},\Lambda^{2})}
+\frac{\xi}{k^{2}} \right\} } & & \nonumber \\
+ \int_{0}^{k^{2}}{ \hspace{-.3cm} d \omega^{2} F(\omega^{2})
\tau(\omega^{2},k^{2}) \left( \frac{\omega^{4}}{k^{4}} - 3
\frac{\omega^{2}}{k^{2}} \right) } & & \nonumber  \\
+ \int_{k^{2}}^{\Lambda^{2}}{ \hspace{-.3cm} d \omega^{2}
F(\omega^{2}) \tau(\omega^{2},k^{2}) \left(
\frac{k^{2}}{\omega^{2}} - 3 \right)
} & = & 0 \,. \nonumber \\
& & \label{transverse vertex RESTRICTION}
\end{eqnarray}
Recall that in the above equation, we have neglected the
contributions of the photon and the scalar bubble diagrams since
they do not contribute to the one loop LLA, Eq.~(\ref{one-loop
sacalar propagator}). Introducing the variable $x$, where
\begin{eqnarray}
x= \frac{\omega^{2}}{k^{2}} & \forall & \omega^{2} \in [0, k^{2}]
\,, \\
x= \frac{k^{2}}{\omega^{2}} & \forall & \omega^{2} \in [k^2,
\Lambda^{2}] \,,
\end{eqnarray}
in Eq.~(\ref{transverse vertex RESTRICTION}), the resulting
restriction can be rewritten as
\begin{equation}
\int_{0}^{1}{ dx \, W(x) } = 0 \,, \label{W restriction}
\end{equation}
with
\begin{eqnarray}
W(x) &=& -6 x \frac{\left(1-x^{\beta}\right)}{x-1}+6 x^{-1}
\frac{\left(1-x^{-\beta}\right)}{x-1} +2 \xi\left( 1 - x^{\beta}
\right) \nonumber \\
& & + \left( x-3 \right) \left( x^{\beta} + x^{-2} \right) h(x)
\,. \label{W definition}
\end{eqnarray}
Note that we have again kept only those terms which contribute to
the LLA. The lower limit $0$ of $x$ integration in Eq.~(\ref{W
restriction}) encodes the fact that we have taken $\Lambda^2
\Rightarrow \infty$. This can be done with impunity as the all
order logarithmic divergence has already been separated out to
construct the MR solution for the wavefunction renormalization
$F$. Moreover, we have introduced the definition
\begin{equation}
h(x) \equiv x k^{2} F(k^{2},\Lambda^{2}) \tau(xk^{2},k^{2}) \,,
\label{H function definition}
\end{equation}
which is a dimensionless function satisfying the property
\begin{equation}
h(x^{-1}) = x^{\beta - 1} h(x) \,, \label{H function property}
\end{equation}
with $\beta = (\xi - 3) / 4\pi$, as prescribed by
Eq.~(\ref{anomalous dimension}). Employing Eq.~(\ref{W
definition}) and the property in Eq.~(\ref{H function property}),
we can write
\begin{eqnarray}
W(x) - W(x^{-1}) &=& 4 \left( x - 1 \right) \left( x^{\beta} +
x^{-2} \right) h(x) \nonumber \\
& & \hspace{-1cm} + 6x \left( 1 - x^{\beta} \right) - 6x^{-1}
\left( 1 -
x^{-\beta} \right) \nonumber \\
& & \hspace{-1cm} + 2 \xi \left[ \left( 1 - x^{\beta} \right) -
\left( 1 - x^{-\beta} \right)  \right] \,. \label{W an W inverse
equation}
\end{eqnarray}
Taking $x=p^{2}/k^{2}$ in~(\ref{W an W inverse equation}), and
using the symmetry $\tau(p^{2},k^{2}) = \tau(k^{2},p^{2})$, it is
straightforward to derive the expression for $\tau(k^{2},p^{2})$
in terms of $W(x)$ and the wavefunction renormalization $F$. On
Wick rotating it back to the Minkowski space, it acquires the
following form:
\begin{eqnarray}
\tau(k^{2},p^{2}) & = & - \frac{3}{2}
\frac{1}{\left(k^{2}-p^{2}\right)} \left[ \frac{1}{F(k^{2})} -
\frac{1}{F(p^{2})}
\right] \nonumber \\
& & \hspace{-2cm} - \frac{\xi}{2}
\frac{1}{\left(k^{2}-p^{2}\right)} \frac{F(k^{2}) +
F(p^{2})}{s(k^{2},p^{2})} \left[ \frac{1}{F(k^{2})} -
\frac{1}{F(p^{2})}
\right] \nonumber \\
& & \hspace{-2cm} + \frac{1}{4} \frac{1}{\left(
k^{2}-p^{2}\right)} \frac{1}{s(k^{2},p^{2})} \left[ W \left(
\frac{k^{2}}{p^{2}}\right) - W \left( \frac{p^{2}}{k^{2}} \right)
\right] \,, \label{tau in terms of W}
\end{eqnarray}
where we have introduced $F(k^{2}) \equiv F(k^{2},\Lambda^{2})$ as
a simplifying notation. We also introduce the definition
\begin{equation}
s(k^{2},p^{2}) = F(k^{2}) \frac{k^{2}}{p^{2}} + F(p^{2})
\frac{p^{2}}{k^{2}} \,. \label{S function definition}
\end{equation}
In the transverse form factor, Eq.~(\ref{tau in terms of W}), the
scalar structure $[1/F(k^{2}) - 1/F(p^{2})]$ appears, as first
reported in spinor QED by Curtis and Pennington in
Ref.~\cite{Curtis:1991fb}. The exact form of the function $W$
remains unknown. Actually, there exists a whole family of
$W$-functions satisfying the integral restriction, Eq.~(\ref{W
restriction}). However, for the sake of simplicity we can choose
the trivial solution $W(x)=0$ for any dimensionless ratio $x$ of
momenta. When substituted in Eq.~(\ref{tau in terms of W}), it
leads to
\begin{eqnarray}
\tau(k^{2},p^{2}) & = & - \frac{3}{2}
\frac{1}{\left(k^{2}-p^{2}\right)} \left[ \frac{1}{F(k^{2})} -
\frac{1}{F(p^{2})}
\right] \nonumber \\
& & \hspace{-2cm} - \frac{\xi}{2}
\frac{1}{\left(k^{2}-p^{2}\right)} \frac{F(k^{2}) +
F(p^{2})}{s(k^{2},p^{2})} \left[ \frac{1}{F(k^{2})} -
\frac{1}{F(p^{2})} \right] \,.  \label{tau with W=0}
\end{eqnarray}
This vertex has already been calculated in one loop perturbation
theory by Bashir \textit{et. al.}, Ref.~\cite{Bashir:2007qq},
using dimensional regularization, in arbitrary gauge $\xi$ and
dimensions $d$.

For the massless case, in dimension $d=4$, they report
\begin{eqnarray}
&& \hspace{-0.6cm} \tau_{BCD}(k^{2},p^{2},q^{2}) =  \nonumber \\
&& \hspace{-0.6cm} \frac{\alpha }{8\pi \Delta
^2}\bigg\{(k^2+p^2-4k\cdot p)\left(k\cdot p
J_0+\ln\left(\frac{q^4}{p^2k^2}\right)\right)
\nonumber \\
&& \hspace{-0.6cm} + \frac{(k^2+p^2)q^2-8p^2k^2}{p^2-
k^2}\ln\left(\frac{k^2}{p^2}\right) \nonumber \\
\nonumber && \hspace{-0.6cm} +  (\xi-1)\left[k^2p^2
J_0+\frac{2[k^2p^2+k\cdot p(k^2+p^2)]}{k^2-
p^2}\right]\ln\left(\frac{p^2}{k^2}\right)
\nonumber \\
&& \hspace{-0.6cm}  +\frac{2k\cdot
p}{k^2-p^2}\left[k^2\ln\left(\frac{q^2}
{p^2}\right)-p^2\ln\left(\frac{q^2}{k^2}\right)\right] \bigg\}
 \,, \label{tau Yajaira }
\end{eqnarray}
where
\begin{equation}
J_{0} = \frac{2}{i\pi^{2}} \int_{M}{ d^{4} \omega
\frac{1}{\omega^{2} \left( p-\omega \right)^{2} \left( k - \omega
\right)^{2}} } \,, \label{J0 definition}
\end{equation}
with $q=k-p$. We now see if our proposal, Eq.~(\ref{tau with
W=0}), fares well against the constraints of this perturbative
form factor, Eq.~(\ref{tau Yajaira }).

\section{\label{sec:PT} Perturbation Theory Constraints}

\begin{figure}[ht]
\includegraphics[width=0.53\textwidth]{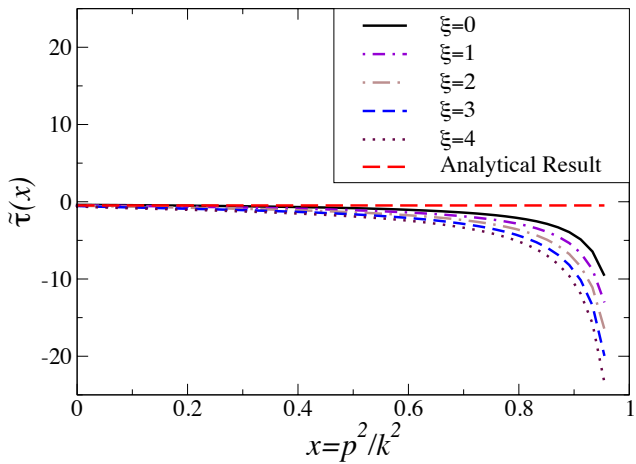}
\caption{\label{fig:-asymp} The analytical result, long dashed
lines representing a constant value given in Eq.~(\ref{tilde-tau})
for the asymptotic transverse form factor $\tilde{\tau}^{\rm
asym}_{\rm BCD}(x)$, agrees with the numerical plot of
$\tilde{\tau}_{\rm BCD}(x)$ obtained from Eq.~(\ref{tau Yajaira })
in the limit $x \rightarrow 0$ for different gauges and an
arbitrarily chosen value of $q^2=-0.7GeV^2$.}
\end{figure}

In order to compare the vertex {\em ansatz}, Eq.~(\ref{tau with
W=0}), based upon multiplicative renormalizability,  against its
one loop perturbative form, Eq.~(\ref{tau Yajaira }), it is
convenient to take the asymptotic limit $k^{2}\gg p^{2}$ of
external momenta in the latter vertex. The resulting $\tau_{\rm
BCD}$ in the LLA is
\begin{equation}
\tau^{\rm asym}_{\rm BCD}(k^{2},p^{2}) \stackrel{k^{2}\gg
p^{2}}{=} -3 \frac{\alpha}{4\pi} \frac{1}{k^{2}} \ln \left(
\frac{k^{2}}{p^{2}} \right) \,.
\label{tau perturbative yajaira}
\end{equation}
Expectedly, it is independent of $q^2$ and hence we drop this
dependence from its argument. Note that this expression is also
independent of the covariant gauge parameter $\xi$. It is unlike
spinor QED where the leading asymptotic vertex is proportional to
$\xi$. For a numerical check, we define
 \begin{eqnarray}
 \tilde{\tau}(x)  = - \frac{k^2 \, \tau(k^{2},x k^{2})}{\alpha \ln x} \,, \label{dimensionless-tau}
 \end{eqnarray}
 where $x=p^2/k^2$ and we have suppressed the $q^2$ dependence for notational
 simplification. Thus:
 \begin{eqnarray}
 \tilde{\tau}^{\rm asym}_{\rm BCD}(x)  = - \frac{3}{4\pi} \,. \label{tilde-tau}
 \end{eqnarray}
 In Fig.~(\ref{fig:-asymp}), we plot
 $\tilde{\tau}^{\rm asym}_{\rm BCD}(x)$ and $\tilde{\tau}_{\rm BCD}(x)$ as
 a function of $x$, the latter for different values of the gauge parameter
 $\xi$ and for a fixed value of $q^2$, chosen arbitrarily. In the
 asymptotic limit, all curves converge to a single value, as
 expected.

Using the perturbative expression, Eq.~(\ref{one-loop sacalar
propagator}), for $F(k^{2})$ in Eq.~(\ref{tau with W=0}), and
taking the asymptotic limit $k^{2}\gg p^{2}$, we have
\begin{equation}
\tau^{\rm asym}(k^{2},p^{2}) \stackrel{k^{2}\gg p^{2}}{=}
\frac{3}{2} \frac{\alpha}{4\pi} \frac{\left( \xi - 3
\right)}{k^{2}} \ln \left( \frac{k^{2}}{p^{2}} \right) \label{tau
perturbative Luis}
\end{equation}
in the LLA. Note that the transverse form factors, Eqs.~(\ref{tau
perturbative yajaira}) and~(\ref{tau perturbative Luis}) have the
functional form $({1}/{k^{2}}) \ln ({k^{2}}/{p^{2}})$.
Furthermore, they are the same in the Feynman gauge ($\xi=1$). In
order for them to be the same in an arbitrary gauge $\xi$, we must
seek a non-trivial $W$-function in Eq.~(\ref{tau in terms of W}),
still satisfying the restriction in Eq~(\ref{W restriction}), so
that the corresponding perturbative vertex is consistent with
Eq.~(\ref{tau perturbative yajaira}) in the asymptotic limit
$k^{2}\gg p^{2}$. Perhaps the simplest such choice for $W$ is
\begin{eqnarray}
W \left( \frac{k^{2}}{p^{2}} \right) &=& \lambda \,
\frac{k^{2}}{p^{2}} \ln \left( \frac{k^{2}}{p^{2}} \right) +
\frac{\lambda}{2} \, \frac{k^{2}}{p^{2}} \,,  \label{W
perturbative} \\ \nonumber
\end{eqnarray}
with $\lambda = - 3\alpha (\xi - 1)/ 2\pi$. In the Feynman gauge
($\xi = 1$) $W=0$, i.e., there is no necessity of a non-trivial
$W$-function since both perturbative vertices, Eqs. (\ref{tau
perturbative yajaira}) and~(\ref{tau perturbative Luis}) are
already the same. Note that the second term in the right hand side
of Eq.~(\ref{W perturbative}) is a convenient term to ensure MR of
the scalar propagator. It drops out in the LLA. Using the variable
$x=k^{2}/p^{2}$ in Eq.~(\ref{W perturbative}), we have
\begin{equation}
W(x) = \lambda \, x  \ln x + \frac{\lambda}{2} \, x \,, \label{W
ansatz}
\end{equation}
so that the restriction in Eq.~(\ref{W restriction}) is trivially
satisfied. Using the choice in Eq.~(\ref{W perturbative}) for $W$
in the vertex, Eq.~(\ref{tau in terms of W}), we can finally
define the transverse form factor as:
\begin{eqnarray}
\tau(k^{2},p^{2}) & = & - \frac{3}{2}
\frac{1}{\left(k^{2}-p^{2}\right)} \left[ \frac{1}{F(k^{2})} -
\frac{1}{F(p^{2})}
\right] \nonumber \\
& & \hspace{-2cm} - \frac{\xi}{2}
\frac{1}{\left(k^{2}-p^{2}\right)} \frac{F(k^{2}) +
F(p^{2})}{s(k^{2},p^{2})}  \left[ \frac{1}{F(k^{2})} -
\frac{1}{F(p^{2})}
\right] \nonumber \\
& & \hspace{-2cm} - \frac{\left( \xi - 1 \right)}{ \left( k^{2} -
p^{2} \right) s(k^{2},p^{2}) } \frac{3\alpha}{8\pi} \left[
\frac{k^{2}}{p^{2}} + \frac{p^{2}}{k^{2}} \right] \ln \left(
\frac{k^{2}}{p^{2}} \right) \,.  \label{tau Final}
\end{eqnarray}
 Note that the Eqs.~(\ref{Ball-Chiu vertex decomposition},\ref{Longitudinal
vertex},\ref{Transverse vertex},\ref{tau Final}) define our full
vertex {\em ansatz}. It ensures the following key features of
sQED:

\begin{table*}
\begin{tabular}{|c|l|c|c|c|}
\hline
& Structure & MR & $\beta$\\
\hline
Bare Vertex & $(k+\omega)^\mu$& No &\\
BC Vertex  & $[ (S^{-1} (\omega)-S^{-1}(k))
(\omega + k)^{\mu}]/(\omega^{2} - k^{2}) $ & No & \\
This work & $\Gamma_{T}^{\mu}(\omega,k)
=\Gamma_{L}^{\mu}(\omega,k)+ \tau(\omega^{2},k^{2},q^{2})[\left(
\omega \cdot q \right) k^{\mu} - \left(k \cdot q
\right)\omega^{\mu}] \label{T mu}$ & Yes & $\alpha(\xi-3)/4\pi$
\\
\hline
\end{tabular} \label{table-1}
\caption{We compare different vertex {\em ans\"{a}tze}: Bare, BC
and our proposal, the last being the only vertex satisfying the
constraints of LKFT and MR. The last column gives the value of the
exponent $\beta$ of the multiplicatively renormalizable
wave-function renormalization in Eq.~(\ref{F unrenormalized lead
log expansion}).} \label{tab_struc}
\end{table*}

\begin{itemize}

   \item It satisfies the WFGTI by
construction,~\cite{Ward:1950xp,Green:1953te,Takahashi:1957xn}.

  \item It guarantees the LKFT property of the
scalar propagator and can be checked by employing it in its SDE.
In other words, it ensures the  multiplicative renormalizability
(MR) of the two point scalar propagator.

    \item It reduces to its one loop perturbation theory Feynman expansion
in the limit of small coupling and asymptotic values of momenta
$k^2 \gg p^2$.

\item It has the same symmetry properties as the bare vertex under
charge conjugation, parity and time reversal, which imply symmetry
under $k \leftrightarrow p$.

    \item It is free of any kinematic singularities when
$k^2 \Rightarrow p^2$, i.e.,
\begin{eqnarray}
   \underset {k^2 \Rightarrow p^2}{\rm lim} \; (k^2 - p^2) \, \tau(k^2,p^2) = 0
   \,.
\end{eqnarray}

\end{itemize}

An important thing to note is that in the {\rm ansatz} for $W$
given in Eq.~(\ref{W ansatz}), MR condition is satisfied
independently of the value of the parameter $\lambda$. Moreover,
$\lambda$ is tied to the anomalous dimensions $\beta$. To the
first order in $\alpha$, we have
 \bea
  \beta = - \frac{\lambda}{6} - \frac{\alpha}{2 \pi} \,.
 \eea
 The NLL and subsequent logs can be obtained
 by writing out:
 \bea
  \beta = \frac{\alpha (\xi-3)}{4 \pi} + c_2 {\cal O}(\alpha^2)
  + c_3 {\cal O}(\alpha^3) + \cdots \,.
 \eea
 Note that the scalar and tensor vertices present in the SDE of
 the scalar propagator, Eq.~(\ref{gap equation}), can start contributing at the NLL and hence
 are required to determine the values of the coefficients $c_i, i\geq
 2$. However, the NLL and constraints from subsequent orders can be absorbed
 in our {\em ansatz} for the effective vector vertex. Practically, this is
 achieved by a new definition for $\lambda$ without
 affecting the MR condition. Therefore, the procedure outlined
 above can easily accommodate the NLL, NNLL and so on. We only
 require $c_i$ for $i=2,3, \cdots$, which are provided by
 increasing orders of perturbation theory, see for example~\cite{Capper:1985nk}.

Note that the kinematic dependence of the vertex on $q^2$ plays no
role asymptotically and the standard analysis proceeds without
reference to it. On the infrared domain, however, the kinematic
dependence on $q^2$ may be important. Our vertex has this pitfall
but its simplicity is reason enough for us to ignore this
dependence.

Finally, in Table~\ref{tab_struc}, we compare different vertex
{\em ans\"{a}tze} as regards the correct behavior of the scalar
propagator under LKFT and MR. Neither the bare vertex nor the BC
vertex yield an MR solution. Our proposal is the only one
satisfying this constraint with the exponent of the wavefunction
renormalization in agreement with the all order LLA in
perturbation theory.

\section{\label{sec:Conc} Conclusions}

   In the massless quenched sQED, we have derived a practical and
easy to implement constraint of multiplicative renormalizability
on the three point scalar-photon vertex. It leads to a family of
these vertices in terms of a constrained dimensionless function
$W(x)$. It has a remarkably simple non perturbative integral
restriction:
\begin{equation}
\int_{0}^{1}{ dx \, W(x) } = 0 \,, \nonumber
\end{equation}
which guarantees the multiplicative renormalizability of the
scalar propagator to all orders in perturbation theory. We further
pin down $W$ through the constraints of one loop perturbation
theory in the asymptotic limit, lack of kinematic singularities
and the imposition of discrete symmetries. Finally, we construct a
simple example ensuring all these key features of the sQED. Though
it is an example from one of the simplest QFTs, it provides a
systematic procedure for constructing a three point function in
terms of the corresponding two point function. This method is
general and can be implemented in a similar manner to unquenched
sQED as well as any other QFT of interest. In this connection, we
would like to comment that an extension to the case of unquenched
sQED is algebraically rather involved. For example for spinor QED,
its unquenched version has been investigated
in~\cite{Kizilersu:2009kg}. It involves the constraints of MR both
on the fermion and photon propagators for massless fermions.
However, the fact remains that in the limit of $n_f \rightarrow
0$, one recuperates the quenched QED results.

Another obvious and straightforward extension of this work is to
apply the same formalism to QCD and constrain the quark-gluon
vertex through the requirements of MR. It will supplement the
earlier works to improve our understanding of this three point
function on
lattice,~\cite{Skullerud:2003qu,Skullerud:2004gp,Kizilersu:2006et},
as well as through continuum
methods,~\cite{Chang:2009zb,Qin:2013mta,Rojas:2013tza}. We
naturally expect the quark-gluon vertex to invoke more W-functions
because the transverse part of this three point vertex is a lot
richer than the one in sQED with eight independent transverse
vectors as compared to only one for the latter. This work is
currently in progress. \\

\noindent {\bf Acknowledgements:} We are grateful to Sixue Qin,
Robert Delbourgo and Alfredo Raya for helpful discussions. This
work was partly supported by CIC, CONACyT and PRODEP grants.


\bibliography{sQEDReferences}

\end{document}